\documentstyle[preprint,epsfig,aps]{revtex}

\def\be{\begin{equation}}
\def\lan{\left\langle}
\def\ran{\right\rangle}
\def\ee{\end{equation}}
\def\barr{\begin{array}}
\def\earr{\end{array}}

\def\nn8{\nonumber\\[10pt]}
\def\l{\left}
\def\r{\right}
\def\dis{\displaystyle}
\def\ed{\end{document}}
\def\cn{{\cal N}}

\def\co{{\cal O}}
\def\cac{{\cal C}}
\def\cach{\hat{{\cal C}}}

\begin{document}

\title{Regularities with random interactions in energy centroids defined by
group symmetries}

\author{V. K. B.  Kota}
\address{Physical Research Laboratory, Ahmedabad 380 009, India}

\maketitle

\begin{abstract}

Regular structures generated by random interactions in energy centroids 
defined over irreducible representations (irreps) of some of the group
symmetries  of the interacting boson models $sd$IBM, $sdg$IBM, $sd$IBM-$T$
and $sd$IBM-$ST$ are studied by deriving trace propagations equations for the
centroids. It is found that, with random interactions, the lowest and highest
group irreps in general carry most of the probability for the corresponding
centroids to be lowest in energy. This generalizes the result known earlier,
via numerical diagonalization, for the more complicated fixed spin ($J$)
centroids where simple trace propagation is not possible.

\end{abstract}




Johnson, Bertsch and Dean in 1998 \cite{Jo-98}, using the nuclear shell
model  found that random two-body interactions lead to ground states, for
even-even  nuclei, having spin $0^+$ with very high probability.  Similarly,
Bijker and Frank \cite{Bi-00} using the interacting boson model with $s$ and
$d$ bosons ($sd$IBM)  showed that random interactions generate vibrational
and rotational structures with high probability. Further studies using the
shell model, fermions in one or two $j$ orbits, $sd$, $sp$ and $sdg$ IBM's,
bosons in a single $\ell$ orbit etc., revealed statistical predominance of
odd-even staggering in binding energies, $0^+$, $2^+$, $4^+$, $\ldots$ yrast
sequence,  regularities in ground states in parity distributions, occupation
numbers and so on; see [3-10] and references therein. Notably, Zelevinsky et
al \cite{Zel-00} introduced the idea of geometric chaos as a basis for the
regularities observed in shell model studies. Similarly,  Zhao et al
\cite{Zh-00} developed a prescription based on sampling of the corners of
the parameter space and  Bijker and Frank \cite{I-mod,Ko-04} employed
mean-field methods. The unexpected results for regularities with random
interactions are reviewed in \cite{regu-1,regu-2}. As Zhao et al stated
\cite{regu-2}: "a more fundamental understanding of the robustness of
$0^+_{g.s.}$ dominance is still out of reach". Therefore, going beyond the
ground states and near yrast levels, energy centroids,  spectral widths and
correlations among them  are also being investigated by several  groups
\cite{Zel-00,Ko-02,Zh-02,Zh-04a,Pa-04} as they are expected to give new
insights into regularities generated by random interactions. For example,
Zhao et al \cite{Zh-02,Zh-04a} initiated the study of energy centroids and
analyzed fixed-$L$ (fixed-$J$, $JT$) centroids in IBM's  (in shell model)
spaces. They found that $L_{min}$ (or $J_{min}$) and $L_{max}$ (or
$J_{max}$) will be lowest with largest probabilities and others appear with
negligible probability. Similarly Papenbrock and Weidenm\"{u}ller
\cite{Pa-04} recently analyzed the structure  of fixed-$J$ spectral widths
for fermions in a single-$j$ shell.

An interesting and important question is the extension of the spin zero
ground state dominance (and also other regular structures seen in shell
model and IBM studies) to group theoretical models with hamiltonians
preserving a symmetry higher than $J$ (or $L$). Similarly one may consider
centroids and variances defined over good or broken symmetry subspaces. They
open a new window to the regularities of  many-body systems in the presence
of random forces.  Initiating work in this direction \cite{Ko-04}, recently
random one plus two-body hamiltonians invariant with respect to $O(\cn_1)
\oplus O(\cn_2)$ symmetry  of a variety of interacting boson models are used
to investigate the probability of occurrence of a given $(\omega_1
\omega_2)$ irreducible representation (irrep)  to be the ground state in
even-even nuclei; $\l[\omega_1\r]$ and $\l[\omega_2\r]$ are symmetric irreps
of $O(\cn_1)$ and $O(\cn_2)$ respectively. It is found that the $0^+$
dominance observed  in ground states of even-even nuclei extends to group 
irreps. The purpose of this paper and others to follow is to go beyond this 
and study regularities, generated by random interactions, in energy
centroids, variances etc. defined over group irreps.  Reported in this rapid
communication are the results of a first analysis of energy centroids  with
examples from $sd$IBM, $sdg$IBM, $sd$IBM-$T$ with the bosons carrying
isospin ($T$) and $sd$IBM-$ST$ with the bosons carrying spin-isospin ($ST$)
degrees of freedom. Before proceeding further, it is important to stress
that energy centroids (also variances) can be calculated as a function of
particle number ($m$) and the quantum numbers labeling the group irreps,
without recourse to the construction of the hamiltonian matrix.  The
principle used here is trace  propagation, a subject introduced in the
context of statistical nuclear spectroscopy by French \cite{Fr-71,Pw-80}.
Readers not interested in the details of group algebra and derivation of
trace propagation equations for the energy centroids (given by Eqs. (5),
(6), (7) and (8)), may leap ahead to the discussion of results starting just
after Eq. (8). 

Let us begin with the spectrum generating algebra (SGA), say  $G_1$, of  a
group theoretical model with all the many particle states in the model 
belong to the irrep  $\Gamma_1$ of $G_1$. For example, the SGA $G_1$ for 
$sd$IBM is $U(6)$. Now the average  of an operator  $\co(k)$ of maximum body
rank $k$ over the irreps $\Gamma_2$ of a subalgebra $G_2$ of $G_1$  ($G_2$
in general denotes a set of subalgebras contained in $G_1$ and $\Gamma_2$
denotes all their irreps) is defined by
\be
\lan \co(k) \ran^{\Gamma_1,\Gamma_2} = \dis\sum_{\beta} \dis\sum_{\alpha \in
\Gamma_2}\;\lan \Gamma_1 \beta \Gamma_2 \alpha \mid \co(k) \mid \Gamma_1 
\beta \Gamma_2 \alpha \ran / \l[\dis\sum_{\beta} \dis\sum_{\alpha \in
\Gamma_2} \lan \Gamma_1 \beta \Gamma_2 \alpha \mid \;1\;
\mid \Gamma_1 \beta \Gamma_2 \alpha \ran \r]
\ee
In Eq. (1), $\beta$ labels the multiple occurrence (multiplicity) of
$\Gamma_2$ in a given $\Gamma_1$ irrep (i.e. in the reduction of $\Gamma_1$
to $\Gamma_2$). Removing  the denominator in Eq. (1) gives the trace over
$(\Gamma_1,\Gamma_2)$ space, i.e. $tr[\co(k)]^{\Gamma_1,\Gamma_2}$.  General
theory for propagation of traces of operators over irreps of group 
symmetries is developed in Refs. \cite{Fr1-79,Fr3-79,Fr4-79,Fr2-79}. In
particular, Quesne \cite{Fr1-79} showed that, for $G_1 \supset G_2$,  trace
propagation over the irreps $\Gamma_1$ and $\Gamma_2$ of $G_1$ and $G_2$
algebras is related to the so-called integrity basis of $G_2$ in $G_1$ which
gives the minimal set of $G_2$ scalars in $G_1$. As discussed in Refs.
\cite{Fr3-79,Fr4-79}, it is seen that in general the multiplicity of
$\Gamma_2$ in a given $\Gamma_1$ irrep results in the  propagation of matrix
of traces $tr[\co(k)]^{\Gamma_1  \Gamma_2;\beta \beta^\prime} =
\sum_\alpha\, \lan \Gamma_1 \beta \Gamma_2 \alpha \mid \co(k) \mid \Gamma_1
\beta^\prime \Gamma_2 \alpha \ran$. However, quite often the  trace of this
trace matrix or its average, as given by Eq. (1), which is important in
applications, may not propagate in a simple manner.  There are approximate
methods for propagating  trace of the trace matrix and they are significant
in particular when the integrity basis contains far too many operators
\cite{Fr4-79}. A very important example here is  fixed-$L$ averages in IBM's
or fixed-$J$ (and $JT$) averages in the shell model. For these, it is not
possible to write a simple propagation  equation in terms of the defining
space averages. On the other hand traces over irreps of group symmetries
(higher than $J$ symmetry) can be propagated in many situations using
Casimir invariants. French and Draayer \cite{Fr2-79} showed that by simple
counting of irreps of  $G_2$ in $G_1$ and the scalars one can construct in
terms of the Casimir invariants of $G_1$ and $G_2$ will  immediately confirm
if propagation via Casimir invariants is possible; in this situation the
integrity basis reduces to Casimir operators of $G_1$ and $G_2$. In this
paper we restrict ourselves to examples in IBM's where this result applies; 
Refs. \cite{Ko-79,Ko-81} give first IBM examples. 

For IBM's the SGA, called $G_1$ above, is $U(\cn)$, with $\cn=6$ for sdIBM,
$15$ for sdgIBM etc. and its irreps $\Gamma_1$ are labeled uniquely by the
boson number $m$ as all $m$ boson states are symmetric with respect to 
$U(\cn)$. Now, consider the average of an operator $\co(k)$  over the irreps
$(m,\Gamma_2)$ with $\Gamma_2$'s being the irreps of a subalgebra $G_2$  of
$U(\cn)$. Say the the number of $(m,\Gamma_2)$'s, called $\Gamma^i$'s
hereafter, for $m \leq k$ is $r$. Also assume that there are $r$ number of
invariants $\cach_i$,  $i=1,2,\ldots,r$ of maximum body rank $k$ constructed
out of the products of $m$ and the Casimir invariants of $G_2$. Then,
for any irrep $\Gamma^0$, clearly   
$\lan \co  \ran^{\Gamma^0} =  \sum_{i=1}^r\; a_i \langle \cach_i
\rangle^{\Gamma^0}$  where $a_i$ are constants. The $a_i$'s can be
determined by assuming that the averages $\lan \co \ran^{\Gamma^j}$ are
known for the irreps  $\Gamma^j$, $j=1,2,\ldots,r$. For example, 
$\Gamma_j$'s can be chosen to be the irreps $(m,\Gamma_2)$'s for $m \leq k$.
With  this, defining the row matrices $[\cac]$ and $[\co_{inp}]$ and the $r
\times r$ matrix $[X]$ as
\be
[\cac] \Leftrightarrow \cac_i = \lan \cach_i \ran^{\Gamma^0}\,,\;\;\;\;
[\co_{inp}] \Leftrightarrow \co_{inp:i}= 
\lan \co \ran^{\Gamma^i}\,,\;\;\;\;
[X] \Leftrightarrow X_{ij} = \lan \cach_j \ran^{\Gamma^i} \;,
\ee
the propagation equation is
\be
\lan \co \ran^{\Gamma} = [\cac]\; [X]^{-1}\;\widetilde{[\co_{inp}]}
\ee
As the eigenvalues of the Casimir invariants of the algebras  $U(\cn)$,
$O(\cn)$ etc. are known, construction of $[\cac]$ and $[X]$ is easy.   In
the reminder of this paper the $H$ is assumed to be $(1+2)$-body. As an
example let us consider $SU(3)$ centroids in $sd$IBM. Here $G_1=U(6)$ and
$G_2=SU(3)$. Simple counting of scalar in terms of the number operator
$\hat{n}$ and the quadratic Casimir operator  ${\hat{\cac}}_2$  and the
cubic Casimir operator ${\hat{\cac}}_3$ of SU(3), confirm that they exhaust
all the scalars needed for propagating $\lan \co (k) \ran^{m,(\lambda
\mu)}$ for any $k$ \cite{Fr4-79,Ko-81}. Note that $(\lambda \mu)$'s denote
$SU(3)$ irreps.  Propagation equation for the energy centroids over $SU(3)$
irreps can be written as $\lan H \ran^{m,(\lambda \mu)} = a_0 + a_1 m + a_2
m^2 + a_3  \cac_2(\lambda \mu)$ where
\be
\cac_2(\lambda \mu) = \lan (\lambda \mu) \alpha \mid {\hat{\cac}}_2 \mid
(\lambda \mu) \alpha \ran = \l[\lambda^2 + \mu^2 + \lambda \mu + 3(\lambda +
\mu)\r]\;. \\
\ee
Using Eqs. (3) and (4), the propagation equation, in terms of the energy 
centroids for $m \leq 2$, is \cite{Ko-81}
\be
\barr{rcl}
\lan H \ran^{m,(\lambda \mu)} & = & \frac{1}{2}\l(2-3m+m^2\r) \lan H \ran^{0,
(00)} + \l(2m-m^2\r) \lan H \ran^{1, (20)} \\
& & + \l[-\frac{5}{6} m + \frac{5}{18} m^2 + \frac{1}{18} 
\cac_2(\lambda \mu)\r] \lan H \ran^{2, (40)} \\
& & + \l[\frac{1}{3} m + \frac{2}{9} m^2 - \frac{1}{18} 
\cac_2(\lambda \mu)\r] \lan H \ran^{2, (02)}
\earr
\ee
Eq. (5) extends easily to the $SU(3)$ limit of $pf$IBM with $U(10)$ SGA but
not to $sdg$, $sdgpf$, etc. IBM's. Now we will derive 3 new propagation 
equations for energy centroids.  

In the $U(\cn) \supset \sum_i\;  \l[U(\cn_i) \supset O(\cn_i)\r] \oplus$
symmetry limits of IBM's, with the bosons carrying angular momenta $\ell_1$,
$\ell_2$, $\ldots$ so that $\cn_i=(2\ell_i+1)$ and $\cn=\sum_i\,\cn_i$, for a
given $i^{th}$ orbit, $U(\cn_i)$ generates number of particles $m_i$ in the
orbit and $O(\cn_i)$ generates the corresponding seniority quantum number
$\omega_i$. The number operators ${\hat{n}}_i$ of $U(\cn_i)$ and the
quadratic Casimir  operators of $O(\cn_i)$ or the corresponding pairing
operators ${\hat{P}}_2(O(\cn_i))$ suffice to give  fixed $\widetilde{m}
\widetilde{\omega} =  (m_1 \omega_1, m_2 \omega_2, \ldots)$ averages of $H$.
Appendix A in Ref. \cite{Ko-00} gives the explicit form of 
${\hat{P}}_2(O(\cn_i))$ for a general situation. Fixed-$\widetilde{m}
\widetilde{\omega}$ centroids of $H$ can be written as $\lan H
\ran^{\widetilde{m} \widetilde{\omega}} = \sum_i\; m_i \epsilon_i +  \sum_{i
\geq j}\;a_{ij}\; m_i (m_j-\delta_{ij}) +  \sum_i\; c_i \lan
{\hat{P}}_2(O(\cn_i)) \ran^{m_i \omega_i}$. Solving for $a_{ij}$'s and
$c_i$'s in terms of the centroids for $m \leq 2$, the final propagation
equation, for IBM's with no internal degrees of freedom, is
\be 
\barr{l} 
\lan H \ran^{\widetilde{m} \widetilde{\omega}} = \dis\sum_i m_i
\epsilon_i +  \dis\sum_{i > j} \overline{V_{ij}}\; m_i m_j + \dis\sum_i
\;\frac{m_i (m_i-1)}{2}\; \lan V \ran^{m_i=2,\omega_i=2} \nn8 
+ \dis\sum_i\;\dis\frac{\lan V \ran^{m_i=2,\omega_i=0} - \lan V
\ran^{m_i=2,\omega_i=2}}{2 \cn_i}\;\l(m_i - \omega_i \r)
\l(m_i+\omega_i+\cn_i-2\r)\;\;; \nn8 
\overline{V_{ij}} = \{\l[\cn_i(\cn_j +
\delta_{ij})\r]/(1+\delta_{ij})\}^{-1} \dis\sum_{L} \; V^{L}_{\ell_i \ell_j
\ell_i \ell_j}  \;(2L+1)\;\;,\nn8 
\lan V \ran^{m_i=2,\omega_i=0} = \lan
(\ell_i \ell_i) L_i=0 \mid V \mid (\ell_i \ell_i) L_i=0\ran\;,\nn8 
\lan V \ran^{m_i=2,\omega_i=2} = \l[\frac{\cn_i (\cn_i +1)}{2}\; 
\overline{V_{ii}} - \lan V \ran^{m_i=2,\omega_i=0} \r]/\l[{\frac{
\cn_i (\cn_i+1)}{2}} -1\r]\;\;. 
\earr 
\ee 
Note that in Eqs. (6), $\epsilon_i$ are energies of the single
particle  levels with angular momentum $\ell_i$ and $V^{L}_{\ell_i \ell_j
\ell_i \ell_j}=\lan (\ell_i \ell_j)L \mid V \mid (\ell_i \ell_j)L \ran$ are
two particle matrix elements of the two-body part of $H$. Also in Eq. (6), 
for $s$ orbit $m_s=2$ and $\omega_s=2$ and there will be no two-boson state
with $\omega_s=0$. Eq. (6) for $sdg$IBM is given first in \cite{Ko-90}, i.e.
for averages over the irreps of the algebras in the chain $U_{sdg}(15) 
\supset U_s(1) \oplus [U_d(5) \supset O_d(5)] \oplus [U_g(9) \supset
O_g(9)]$. Similarly Eq. (6) gives $H$ averages over the irreps of $U_{sd}(6)
\supset U_d(5) \supset O_d(5)$ of $sd$IBM, $U_{sdpf}(16) \supset [U_d(5)
\supset O_d(5)] \oplus [U_p(3) \supset O_p(3)] \oplus [U_f(7) \supset
O_f(7)]$ of $sdpf$IBM etc. Moreover this extends easily (this will be
discussed elsewhere) to IBM's with internal degrees of freedom. Let us add 
that it is also possible to write down propagation equations for the variances 
$\langle [H - \langle H \rangle^{ \widetilde{m} \widetilde{\omega}}]^2 
\rangle^{\widetilde{m} \widetilde{\omega}}$ using the results in 
\cite{Ko-79,Qs-78}. 

In IBM-$T$ with $U(3\cn) \supset U(\cn) \otimes [SU_T(3) \supset O_T(3)]$ 
where $U(\cn)$ gives the spatial part (for $sd$, $sdg$, $sdpf$ etc.) and
$O_T(3)$ generating  isospin \cite{El-80}, it is possible to propagate the
centroids $\lan H \ran^{m, \{f\}, T}  \equiv \lan H \ran^{m,(\lambda \mu),
T}$. Note that the  $U(\cn)$ irreps are labeled by $\{f\}=\{f_1,f_2,f_3\}$
where $f_1 \geq f_2 \geq f_3 \geq 0$ and $m=f_1+f_2+f_3$. The corresponding
$SU_T(3)$ irreps are $(\lambda,\mu) = (f_1-f_2,f_2-f_3)$. The $SU_T(3)$ to
$O_T(3)$ reductions follow from the formulas given by Elliott
\cite{El-58,Ko-98}.  The scalars $1$, $\hat{n}$, $\hat{n}^2$, $\cach_2(SU_T(3)
)$ and ${\hat{T}}^2$ and the energy centroids for $m \leq 2$, via Eqs. (2), 
(3) and (4), give
\be
\barr{rcl}
\lan H \ran^{m,(\lambda \mu), T} & = &
\l[1-\frac{3}{2}m + \frac{m^2}{2}\r]
\lan H \ran^{0, (00), 0} + \l[2m-m^2\r] \lan H \ran^{1, (10), 1} \\
& & + \l[-\frac{1}{6}m + \frac{1}{18}m^2 + \frac{1}{9} \cac_2(\lambda \mu)
-\frac{1}{6} T(T+1) \r] \lan H \ran^{2, (20), 0} \\
& & + \l[-\frac{5}{6}m + \frac{5}{18}m^2 + \frac{1}{18} \cac_2(\lambda \mu)
+\frac{1}{6} T(T+1) \r] \lan H \ran^{2, (20), 2} \\
& & + \l[\frac{1}{2}m + \frac{1}{6}m^2 - \frac{1}{6} \cac_2(\lambda \mu)
\r] \lan H \ran^{2, (01), 1} \;\;.
\earr
\ee
For $sd$IBM-$T$, starting with the general hamiltonian given in Appendix-A of
\cite{Ko-98} which contains the $s$ and $d$ boson energies and 17
two-particle matrix elements $V^{L,t}_{\ell_l \ell_2 \ell_3 \ell_4}$, it
is easy to write down $\lan H \ran^{m,(\lambda \mu), T}$
for $m \leq 2$; for $m =2$ the two-boson isospins $t$ uniquely
define the corresponding $SU_T(3)$ irreps.  Thus Eq. (7) for $\lan H
\ran^{m,(\lambda \mu), T}$ is easy to apply for any $m$. 

In IBM-$ST$, a group chain of interest is \cite{El-81}  $U(6\cn) \supset
U(\cn) \otimes \l[ SU_{ST}(6) \supset O_{ST}(6)\r]$ with $U(\cn)$ generating
the spatial part and $SU_{ST}(6)$ [or $U_{ST}(6)$] generating spin-isospin
part; note that the Wigner's spin-isospin super-multiplet algebra 
$SU_{ST}(4)$ is isomorphic to $O_{ST}(6)$. Just as before, it is possible to
propagate the centroids $\lan H\ran^{m,\{f\}, [\sigma]}$.  Here $\{f\}$'s are
the irreps of $U(\cn)$ or equivalently $U_{ST}(6)$ and
$\{f\}=\{f_1,f_2,f_3,f_4,f_5,f_6\}$  where $\sum_i\,f_i =m$ and $f_i \geq
f_{i+1} \geq 0$. The $O_{ST}(6)$ irreps are labeled by $[\sigma]=[\sigma_1,
\sigma_2, \sigma_3]$ and the $\{f\}$ to $[\sigma]$ reductions, needed for the
results discussed ahead, follow from the  analytical formulas given in
\cite{Ko-98} and the tabulations in \cite{Wy-70}. Eqs. (2) and (3) give,
using the quadratic Casimir invariants ${\hat{\cac}}_2$'s of $U_{ST}(6)$ and
$O_{ST}(6)$, 
\be
\barr{rcl} 
\lan H \ran^{m,\{f\}, [\sigma]} & = & \l[1-\frac{3}{2}m +
\frac{m^2}{2}\r] \lan H \ran^{0, \{0\}, [0]} + \l[2m-m^2\r] \lan H \ran^{1,
\{1\}, [1]} \\ 
& & +\;\l[-\frac{5}{3}m + \frac{1}{4}m^2 + \frac{1}{6} \cac_2(\{f\}) 
+\frac{1}{12} \cac_2([\sigma]) \r] \lan H \ran^{2, \{2\}, [2]}  \\ 
& & +\;\l[-\frac{1}{12}m + \frac{1}{12} \cac_2(\{f\})
-\frac{1}{12} \cac_2([\sigma]) \r] \lan H \ran^{2, \{2\},[0]} \\  
& & +\;\l[\frac{5}{4}m + \frac{1}{4}m^2 - \frac{1}{4} \cac_2(
\{f\})\r] \lan H \ran^{2, \{1^2\}, [1^2]} \;; 
\earr 
\ee
where $\cac_2(\{f\})=\lan \hat{\cac}_2(U_{ST}(6))\ran^{\{f\}} = 
\dis\sum_{i=1}^6\,f_i \l(f_i+7-2i\r)$ and $\cac_2([\sigma]) = 
\lan \hat{\cac}_2(O_{ST}(6))\ran^{[\sigma]} = \dis\sum_{i=1}^3\,
\sigma_i \l(\sigma_i+6-2i\r)$.
Diagonalizing $\hat{\cac}_2(O_{ST}(6))$ in the $\l| (\ell_1 \ell_2)LST\ran$ 
basis and applying the resulting unitary transformation to the $H$ matrix 
in this basis will give the input averages in Eq. (8).

Now we will apply Eqs. (5)-(8) to study regularities generated by random
interactions in energy centroids. In all the calculations used are independent 
Gaussian random variables with zero center and unit variance and a 1000 
member ensemble. We begin with the simplest example of $sd$IBM centroids. The
highest  $SU(3)$ irrep for a given $m$ is $(2m,0)$  and Eq. (5) gives, $\lan
H \ran^{m,(\lambda \mu)} - \lan H \ran^{m,(2m,0)} = [\cac_2(\lambda \mu) -
\cac_2(2m,0)]\;\Delta/18$ with $\Delta= \lan H \ran^{2,(40)} - \lan H
\ran^{2,(02)}$. Therefore the probability of finding $\Delta$ to be positive
or negative will simply give the probability for finding the highest or
lowest $m$ particle $SU(3)$ irrep to be lowest in energy. With the
two-particle matrix elements chosen to be Gaussian variables (with zero
center and unit variance), $\Delta$ itself will be a Gaussian variable with
zero center.  For $m=3k$, $3k+1$ and $3k+2$, $k$ being a positive integer,
the lowest $SU(3)$ irreps are $(00)$, $(20)$ and $(02)$ respectively.  They
will be lowest in energy with $50$\% and the $(2m,0)$ irrep will be lowest in
energy with $50$\% probability. Thus, it is easy to understand the
regularities in centroids defined over fixed  $SU(3)$ irreps in $sd$IBM with
one plus two-body hamiltonians, without constructing the many boson
hamiltonian matrix but just by using the  propagation equation (5). 

\begin{figure}
\epsfig{width = 4in, height = 3in, figure = 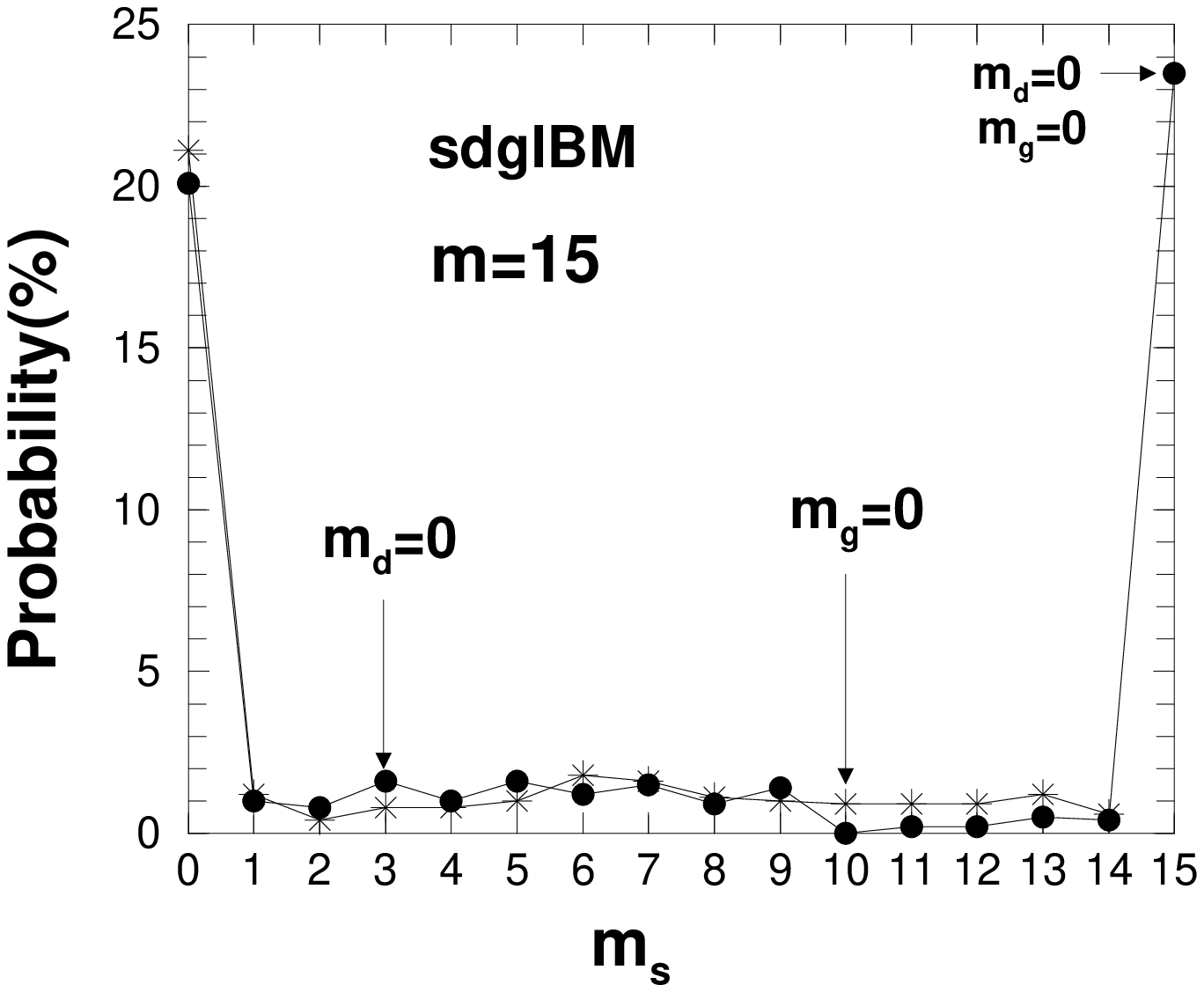}
\vskip 0.5cm
\caption{Probabilities for $sdg$IBM fixed-$(m_s,m_d,v_d,m_g,v_g)$ centroid 
energies to be lowest in energy vs $m_s$ for a system of 15 bosons ($m=15$). 
For each $m_s$, the  probability shown is the sum of the probabilities for
the irreps with the seniority quantum  number lowest ($v_\ell =
\pi(m_\ell)$) and highest ($v_\ell=m_\ell$). Filled circles and stars are
for configurations with $m_d=0$ and $m_g=0$ respectively; they are joined by
lines to guide the eye. Note that for  $m_s=15$ both $m_d=0$ and $m_g=0$.}
\label{sdg}
\end{figure}

In $sdg$IBM, regularities in fixed-$(m_s,m_d,v_d,m_g,v_g)$ centroids are
studied using the propagation Eq. (6). Choosing the 3 single particle
energies $(\epsilon_s, \epsilon_d, \epsilon_g)$ and the 16 diagonal
two-particle matrix elements $V^L_{\ell_1 \ell_2 \ell_1 \ell_2}$, with
$\ell_i=0,2$ and 4 to be Gaussian variables, the probability for the
centroid of a  given $(m_s,m_d,v_d,m_g,v_g)$ configuration to be lowest is
calculated for $m=6-25$ and the results are shown in Fig. \ref{sdg} for
$m=15$. To maintain proper scaling, the $\epsilon$'s are divided by $m$ and
the $V^L$ by $m(m-1)$ just as in \cite{Bi-00}.  For the discussion of the
results we define $\pi(x)$ such that $\pi(x)=0$ for $x$ even and $\pi(x)=1$
for $x$ odd.  It is seen from Fig. 1,  and also valid for any $m$, that the
configurations $(m_s,m_d=v_d=m-m_s,m_g=v_g=0)$,
$(m_s,m_d=m-m_s,v_d=\pi(m_d),m_g=v_g=0)$, $(m_s,m_d=v_d=0,m_g=v_g=m-m_s)$
and $(m_s,m_d=v_d=0,m_g=m-m_s, v_g=\pi(m_g))$ exhaust about 91\%
probability. Moreover, the configurations with $m_s=m_d=0$ carry $\sim
20$\%, $m_s=m_g=0$ carry $\sim 21$\%, $m_s=m$ carries $\sim 24$\% and $m_s
\neq 0$ but $m_d=0$ or $m_g=0$ carry $\sim 26$\% probability.  Thus the
$m_s=m$ configuration and the four configurations  with $m_s=0$ are most
probable to be lowest in energy. However, other  configurations with $m_s
\neq 0, m$ (they are 49 out of 1195,  configurations in the $m=15$ example)
give  non-negligible probability for being lowest. Thus about $\sim 4$\% of
the $(m_s,m_d,v_d,m_g,v_g)$  configurations will have probability to be
lowest with random interactions. 

\begin{figure}
\epsfig{width = 4in, height = 3in, figure = 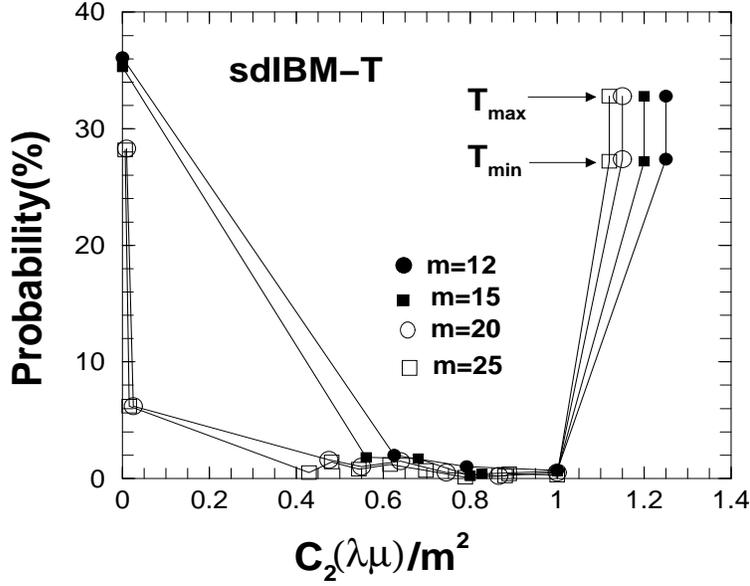}
\vskip 0.5cm
\caption{Probabilities for the $sd$IBM-$T$'s $(\lambda \mu)T$  centroid
energies to be lowest in energy vs $\cac_2(\lambda \mu)/m^2$ for boson
systems with $m=12,15,20$ and $25$.  Except for the highest $(\lambda \mu)$,
for all other $(\lambda \mu)$'s shown in the figure, the probabilities are
for $T_{max}$ if $\lambda \neq 0$ and  $\mu \neq 0$ and they are for
$T_{min}$ if $\lambda=0$ or $\mu=0$. For the irreps not shown in the
figure, the probability is $< 0.1$\%. All the points for a given $m$ are
joined by lines to guide the eye.}
\label{ibm3}
\end{figure}

For $sd$IBM-$T$, it is easily seen from Eq. (7) that the one-body part of
$H$ will not play any role in the study of fixed-$(\lambda, \mu)T$
centroids. Choosing $V^{L,t}_{\ell_1 \ell_2 \ell_1 \ell_2}$'s to be  
Gaussian  variables,  the centroids are generated, using Eq. (7), for
$m=10-25$ and for all allowed $(\lambda, \mu)T$.   Some typical results for
the regularities are shown in Fig. \ref{ibm3}. Firstly, for a given $m$ the
highest $SU_T(3)$ irrep is  $(m,0)$ with $T_{max}=m$ and $T_{min} =\pi(m)$.
For $m=3k$, $3k+1$ and $3k+2$, with $k$ being a positive integer, the lowest
$SU_T(3)$ irreps are $(00)$, $(10)$ and $(01)$ with $T=0$, $1$ and $1$
respectively; for the later two situations the next lowest irreps are $(02)$
and $(20)$ respectively with $T_{min}=0$. For $m=3k$, it is seen from Fig. 2
that the  lowest $SU_T(3)$ irrep's centroid (here $T$ is unique) is lowest
with  $\sim 35$\% probability. Similarly the highest irreps centroid is
lowest with $\sim 60$\% probability and this splits into $\sim 30$\% each
for the lowest and highest $T$'s. For $m=3k+1$ and $3k+2$, the probability
for the centroid of the highest irrep  to be lowest in energy is same as for
$m=3k$. However for the centroid of the lowest irrep, the probability is
$\sim 29$\% and the next lowest irrep appears with $\sim 6$\%. Thus in
general the centroids of the highest and the lowest (for $m=3k+1$ and
$3k+2$, the lowest two) $SU_T(3)$ irreps exhaust about 95\% of the
probability for being lowest in energy.  As the two particle centroids
$X^t=\lan H \ran^{m=2,t}$ are linear  combinations of  $V$'s, it can be seen
that they themselves are Gaussian variables.  Note that $\lan H
\ran^{m,(\lambda \mu)T} -\lan H \ran^{m,(m,0)m}=[\cac_2(\lambda \mu) -
\cac_2(m,0)]\;\Delta_1 + [T(T+1)-m(m+1)]\;\Delta_2$ where $\Delta_1 =
\frac{1}{9}X^0 +\frac{1}{18} X^2 -\frac{1}{6} X^1$ and $\Delta_2 =
\frac{1}{6}(X^2-X^0)$. Calculations with $X^t$'s taken as Gaussian variables
with same variance (actually the variance of $X^0$ and $X^2$ are same and
that of $X^1$ is $\sim 20$\% higher) are carried out and it is seen that
they give almost same results as in Fig. 2.  

\begin{figure}
\epsfig{width = 4in, height = 3in, figure = 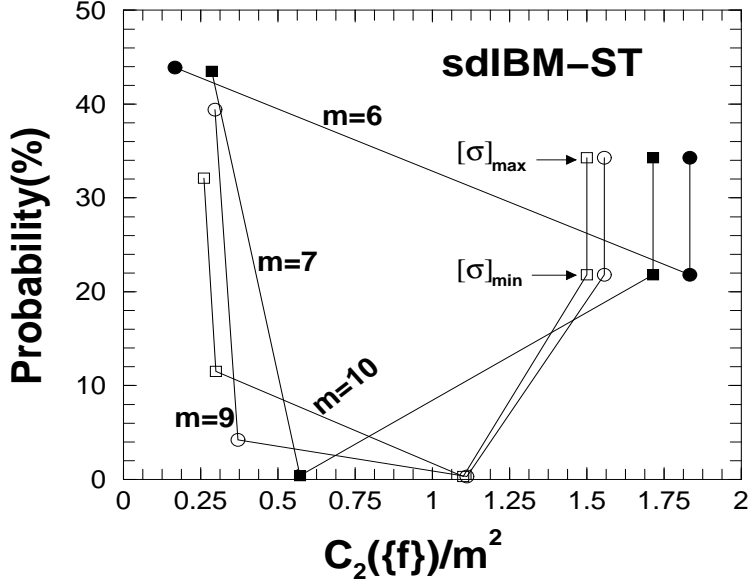}
\vskip 0.5cm
\caption{Probabilities for the $sd$IBM-$ST$'s $\{f\} [\sigma]$ centroid 
energies to be lowest in energy vs $\cac_2(\{f\})/m^2$ for boson systems with
$m=6,7,9$ and $10$. For the structure of the irreps with probability $>2$\%, 
see text. For $m=7,9$ and $10$ there is one additional irrep with $\sim 0.3$\% 
probability. For the irreps not shown in the figure, the probability is 
$< 0.1$\%. All the points for a given $m$ are joined by lines to guide 
the eye.}
\label{ibm4}
\end{figure}

For $sd$IBM-$ST$, as seen from Eq. (8), the energy centroids $\lan H
\ran^{m,\{f\}, [\sigma]}$  are determined by the 2-particle averages $\lan H
\ran^{2, \{2\}, [2]}$, $\lan H \ran^{2, \{2\}, [0]}$ and $\lan H \ran^{2,
\{1^2\}, [1^2]}$ and they are linear combinations of the two particle matrix
elements $V^{LST}$ in the $\l| (\ell_1 \ell_2)LST\ran$ basis. Instead of
choosing $V^{LST}$ to be Gaussian variables, we have chosen,  using the
result found in the $sd$IBM-$T$ examples, the three 2-particle averages to
be Gaussian variables. Using this, the probabilities are  calculated for
various $m$ values and some of the results are shown in Fig. \ref{ibm4}.
Firstly for a given $m$ the highest $\{f\}$ is $\{m\}$. The corresponding
highest and lowest $[\sigma]$ are $[m]$ and $[\pi(m)]$. For all $m$ the
centroid of the highest  $U_{ST}(6)$ irrep is lowest with $\sim 56$\%
probability and this splits into $\sim 34$\% and $\sim 22$\% for the highest
and lowest $O_{ST}6)$ irreps. For $m=6k$, $6k \pm 1$, $6k \pm 2$ and $6k+3$,
with $k$ a positive integer, the lowest $U_{ST}(6)$ irreps are those that
can be reduced to the irreps $\{0\}$, $\{1\}$, $(\{2\}, \{1^2\})$ and
$(\{1^3\}, \{21\})$ respectively. These irreps with the corresponding
lowest $[\sigma]$ are lowest with probability $\sim 43$\%. 

In conclusion, with random interactions, the lowest and highest group irreps
(i.e. irreps of $G_2$ in $G_1 \supset G_2$) carry most of the probability
for the corresponding centroids to be lowest in energy. With the inclusion
of a subalgebra ($G_1 \supset G_2 \supset G_3$), these probabilities split
into the probabilities for the corresponding lowest and highest irreps of
the subalgebra. This is indeed the situation for all the examples discussed
in this paper. Continuing with the process of embedding  subalgebras, the
$O(3)$ algebra generating $L$ can be reached (with generalization for
systems with $LT$, $LST$ or $JT$). Then clearly the energy centroids of
highest and lowest $L$'s should be most probable and this is found to be
true numerically in \cite{Zh-02,Zh-04a}. An important aspect of the energy
centroids is that they propagate via Casimir invariants in many situations.
New propagation equations are derived in this paper (Eqs. (6), (7) and (8)).
In fact there are many other situations  where such equations can be
derived; an example is for the centroids over the irreps $[m_{sd}
(\lambda_{sd} \mu_{sd}); m_{pf} (\lambda_{pf} \mu_{pf})]$ of $[U_{sd}(6)
\supset SU_{sd}(3)] \oplus [U_{pf}(10) \supset SU_{pf}(3)]$ algebra of
$sdpf$IBM \cite{sdpf}. These will be discussed in a longer paper along with 
extensions of the present work to spectral variances and also to shell model
symmetries. Finally, an important observation is that the propagators carry
information about $G_1 \supset G_2$ geometry (i.e. $G_1 \supset G_2$ reduced
Wigner coefficients and $G_2$ Racah coefficients) and thus it is plausible
that propagation equations may be useful in quantifying  geometric chaos.
This is being investigated and it should be remarked that only  recently the
role of Wigner-Racah algebra in two-body random matrix ensembles  is
established \cite{Ko-05}.

Thanks are due to Y.M. Zhao for useful correspondence and for making
available Ref. \cite{Zh-04a} before it is submitted for publication.

\newpage

\newpage

\newpage

\ed